\documentclass[twocolumn,secnumarabic,amssymb, nobibnotes, aps, prb]{revtex4-1}

\usepackage{amsfonts}
\usepackage{comment}
\usepackage{relsize}
\usepackage{amsmath}
\usepackage{amssymb}
\usepackage{graphicx}
\usepackage{epstopdf}
\usepackage[dvipsnames]{xcolor}
\usepackage[toc,page]{appendix}
\usepackage{float}%
\usepackage{comment}
\begin{document}
	\begin{abstract}
		The low energy Dirac and Weyl spectra are allowed to violate the Lorentz symmetry and thereby have a tilted energy dispersion. The tilt in the energy dispersion induces a Hall voltage in the plane spanned by the electric field and the tilt velocity. In the presence of a magnetic field the planar Hall conductivity and resistivity show Shubnikov de Haas oscillations. The oscillations in the planar Hall effect can become a fingerprint to spot the anomalous transport in Dirac and Weyl semimetals.   
		\begin{description}
			\item[Usage]
			.
			\item[PACS numbers]
			May be entered using the \verb+\pacs{#1}+ command.
			\item[Structure]
			You may use the \texttt{description} environment to structure your abstract;
			use the optional argument of the \verb+\item+ command to give the category of each item. 
		\end{description}
	\end{abstract}
	\title{Exploiting the violation of Lorentz symmetry for the planar Hall effect}%
	
	\author{Muhammad Imran and Selman Hershfield}%
	\affiliation{Department of Physics, University of Florida, Gainesville, Florida 32611,USA}
	\date{\today}%
	\maketitle
	\section*{Introduction}
The Dirac equation predicts three elementary fermions. These are well known by the names of Dirac, Majorana, and Weyl fermions. Two of them, the Majorana and Weyl fermions, have not been experimentally observed as a fundamental particle. However, this family of fermions has been experimentally detected in the low energy physics.\cite{RefI1,RefI2,RefI3} 

The speed of electrons and holes in the energy bands of solids is always less than the speed of light. Therefore it is not a surprise that these fermions violate the fundamental Lorentz symmetry, which is not possible for the elementary particles.\cite{RefI4} There are two types of fermions that violate the Lorentz symmetry in the low energy physics.\cite{RefI5} This classification can be explained by the energy dispersion equation. The constant energy surfaces for Dirac and Weyl fermions is a spheroid. With type-I tilt the constant energy surfaces are ellipsoids, and with type-II tilt constant energy surfaces are hyperboloids. This classification is also made from the boundary where the conduction and valance bands meet. The type-I Dirac and Weyl fermions bands converge to a single point. The type-II Dirac and Weyl fermions bands meet and form a non-closed isoenergic orbit. The Weyl fermions emerge from the Dirac fermions provided any or both of the two the time reversal and the inversion symmetries are absent.\cite{RefI32} Both types of Weyl fermions are topologically protected excitations. The type-I and type-II Dirac and Weyl fermions have been observed in the angle resolved photo emission spectroscopy and the magnetotransport measurements.\cite{RefI4,RefI6}

The tilt in energy dispersion of the Dirac and Weyl fermions is parameterized by the tilt velocity. In general, the tilt velocity modifies the energy spectrum. This propagates in the energy density of states and the magnetotransport properties. It should be noted that a strong magnetic field can not quantize the motion of charged particles in a hyperboloid surface. Therefore no quantized solutions exists for magnetic field applied perpendicular to the direction of type-II tilt velocity.\cite{RefI5,RefI15} In this study we show that the tilt velocity mixes different elements of the conductivity matrix. This induces the Hall voltage in the plane formed by the perpendicular electric and magnetic fields, provided the tilt velocity makes finite angle in this plane. Although the planar Hall voltage  survives in the absence of the magnetic field, the magnetic field induces quantization signatures, the Shubnikov de Haas oscillations. 

The semiclassical theory of the planar Hall effect in the three dimensional(3D) Weyl semimetals has been studied,\cite{RefI33,RefI34} and experimentally observed.\cite{RefI12} Those semiclassical theories of the planar Hall effect rely on the chiral anomaly, a Berry curvature effect in magnetotransport. According to this theory for an anomalous semimetal longitudinal magnetoresistivity decreases quadratically in the direction of magnetic field. Although there are experimental results that observe the planar Hall effect, the anomalous magnetoresistivity expected with the chiral anomaly was not observed.\cite{RefI38} In general the planar Hall effect can arise even without magnetic field.\cite{RefI36} We are studying the planar Hall effect both in the quantum and semiclassical regions of magnetic field. In this study we show the origin of the planar Hall effect in the Dirac and Weyl semimetals can also be a tilted energy dispersion even without chiral anomaly. 

We shall discuss the theoretical formulation and model Hamiltonian for a general 3D Dirac and Weyl fermions in section-II, the numerical results and discussions are presented in section -III, and finally conclusions are given in section- IV.

 \section*{Theoretical formulation}
The Hamiltonian for a tilted Dirac and Weyl semimetals is
\begin{equation}
H=\chi(v_{T}^{i}p_{i}\sigma^{0}+v_{F}p_{i}\sigma^{i}),
\end{equation}
where $\sigma^{i}$ and $p_{i}$ denotes the $x,y$, and $z$ components of the Pauli matrix and the momentum. Summation is assumed in the repeated indexes. $\chi=\pm$ is the chirality of the Weyl fermions, and $v_{F}$ is Fermi velocity. The eigenvalues of this Hamiltonian are $\epsilon(s,p)=\chi(v_{T} p\cos\theta+svp)$, where $s=\pm$ denotes the band index and $\theta$ is the angle between the tilt velocity and the momentum. 
The energy density of states for a tilted Dirac and Weyl spectrum has a parabolic energy dependence in the normalized multiplying constant.
\begin{equation}
\begin{split}
\frac{D(E)}{D_{0}(E)}=\begin{bmatrix}
\frac{g_{1}}{(1-(\frac{v_{T}}{v_{F}})^{2})^{2}} & v_{F}>v_{T}&\textrm{Type-I}\; \\
\frac{g_{2}}{(1-(\frac{v_{F}}{v_{T}})^{2})^{2}}\frac{v_{F}^{3}}{2v_{T}^{3}}(1+(\frac{v_{F}}{v_{T}})^{2}) & v_{T}>v_{F} &  \textrm{Type-II}
\end{bmatrix}
\end{split}
\end{equation}
In the above formula the tilt is assumed along $z$-axis, $\vec{v}_{T}=v_{T}^{z}\hat{z}$. Here $D_{0}(E)=\mathlarger{\frac{E^{2}}{2\pi^{2}\hbar^{3}v_{F}^{3}}}$ shows the energy density of states for the Dirac and Weyl spectra without any tilt velocity.\cite{RefI5}
  The symbol $g_{1(2)}$ refers to the number of type-I(II) Dirac and Weyl nodes. The energy density of states diverges for $\frac{v_{T}}{v_{F}}\rightarrow1$(see Ref. [15] for discussions on this topic).

 We assume the $z$-axis is along the magnetic field $\vec{B}=B_{z}\hat{z}$.  $v_{T}^{z}$ is the tilt in the direction of magnetic field, and $v_{T}^{x}$ in the direction perpendicular to the magnetic field. We take $v_{T}^{x}$ to be nonzero only for type-I Dirac and Weyl semimetals because one can not quantize the Hamiltonian with our approach for type-II semimetals. Thus the Hamiltonian including tilt is
\begin{equation}
\begin{split}
H=\chi(v_{T}^{x}p_{x}+v_{T}^{z}p_{z}+v_{F}\vec{\sigma}\cdot\vec{p}).
\end{split}
\end{equation}
The energy eigenvalues and energy eigenstates are found after transforming the above Hamiltonian by hyperbolic transformation in the $\hat{x}$ direction,
\begin{equation}
(\epsilon-H)|\psi\rangle\rightarrow\bar{\epsilon}-\bar{H})|\bar{\psi}\rangle.
\end{equation}
 The bar on top of the operators and the eigenstates show hyperbolic transformation taken along the $\hat{x}$ axis.
\begin{eqnarray}
\bar{O}&=&N^{2}\exp[\sigma_{x}\frac{\theta}{2}]O\exp[\sigma_{x}\frac{\theta}{2}]\\
|\bar{\psi}\rangle&=&\frac{1}{N}\exp[-\sigma_{x}\frac{\theta}{2}]|\psi\rangle,
\end{eqnarray}
Here $N$ is the normalization constant of the hyperbolic transformation.
For convenience we use the Landau gauge for solving this problem: $p_{y}\rightarrow p_{y}-eA_{y}$, with $A_{y}=B_{z}x$.
The energy eigenvalues are
\begin{eqnarray}
\epsilon^{s}_{n}(p_{z})=\chi(v_{T}^{z}p_{z}+\frac{s}{\gamma}\sqrt{v_{F}^{2}p_{z}^{2}+\frac{2n\hbar^{2}v_{F}^{2}}{\gamma l_{B}^{2}}}).
\end{eqnarray}
Here $s=\pm$, and $\gamma^{-1}=\sqrt{1-(\frac{{v_{T}^{x}}}{v_{F}})^{2}}$. The corresponding energy eigenstates are
\begin{eqnarray}
\psi^{+}_{n}&=&\frac{1}{\sqrt{2}}\begin{bmatrix}
a_{n}\phi_{n}(x-x_{0})\\
-ib_{n}\phi_{n-1}(x-x_{0})
\end{bmatrix}\\
\psi^{-}_{n}&=&\frac{1}{\sqrt{2}}\begin{bmatrix}
b_{n}\phi_{n}(x-x_{0})\\
ia_{n}\phi_{n-1}(x-x_{0})
\end{bmatrix}.
\end{eqnarray}
The energy eigenstates are shifted by the displacement $x_{0}=\frac{p_{y}l_{B}^{2}}{\hbar}$. The symbol $l_{B}$ is used for the magnetic length, and it is related with the magnetic momentum $p_{B}$ by $p_{B}=\frac{\sqrt{2}\hbar}{l_{B}}$.
The normalization constants are
\begin{eqnarray}
a_{n}&=&\sqrt{1+\frac{v_{F}p_{z}}{\gamma\sqrt{v_{F}^{2}p_{z}^{2}+\frac{2n\hbar^{2}v_{F}^{2}}{\gamma l_{B}^{2}}}}}\\
b_{n}&=&\sqrt{1-\frac{v_{F}p_{z}}{\gamma\sqrt{v_{F}^{2}p_{z}^{2}+\frac{2n\hbar^{2}v_{F}^{2}}{\gamma l_{B}^{2}}}}}.
\end{eqnarray}
\begin{figure}[b!]
	\centering
	\includegraphics[width=80mm]{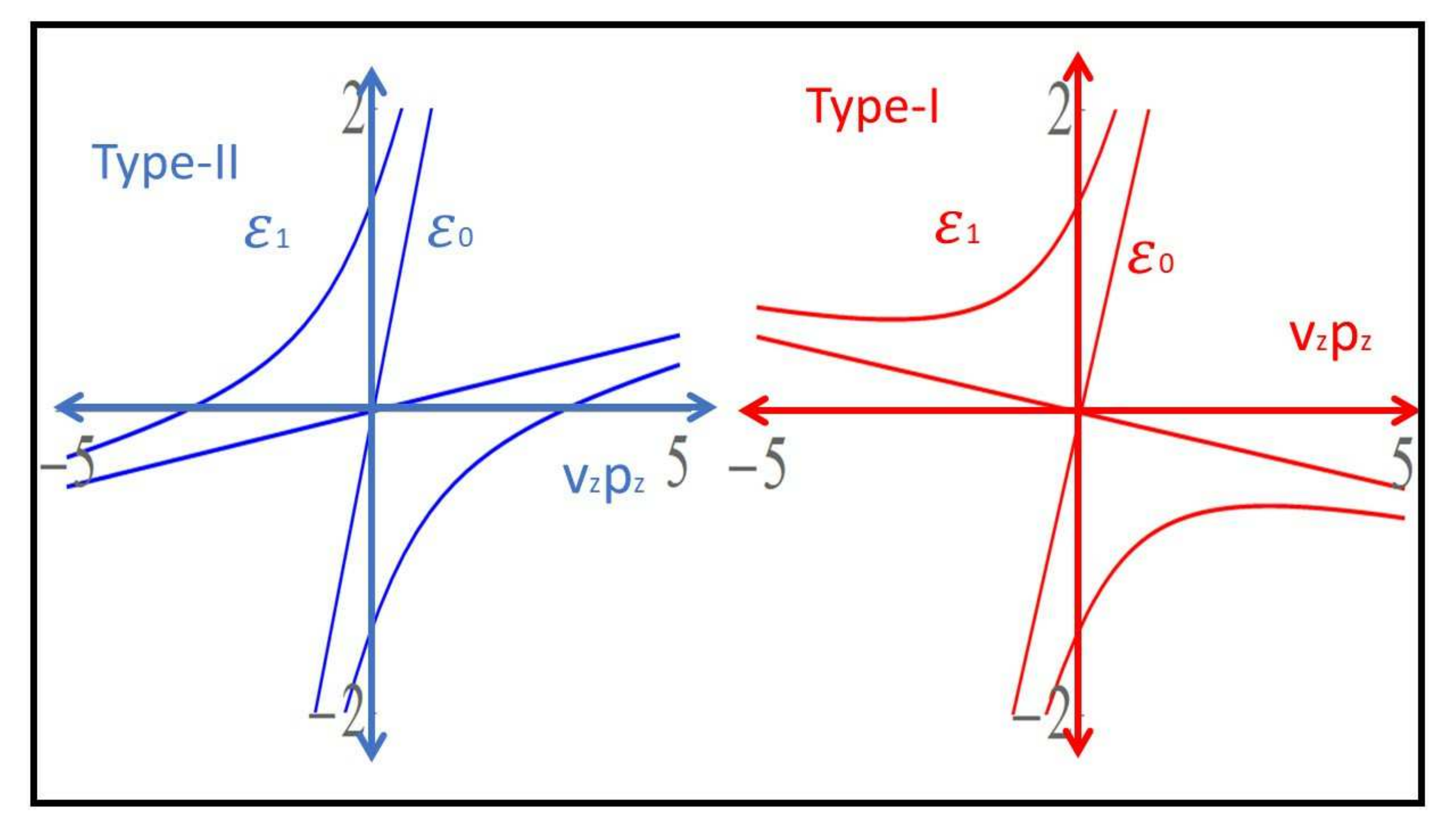}
	\caption{The energy spectra of Weyl gases. We assume the Lorentz factor $\gamma=1$, the type-I tilt velocity $\frac{v_{T}^{z}}{v_{F}}=0.9$, and the type-II tilt velocity $\frac{v_{T}^{z}}{v_{F}}=1.1$.  \label{Pic1}}
\end{figure}
 Ground state, $n=0$, is the gapless state with eigenvalue
 \begin{eqnarray}
\epsilon^{s}_{n=0}(p_{z})&=&\chi(v_{T}^{z}+s (sgn(p_{z})v_{z}))p_{z}.
 \end{eqnarray}
 The energy spectra of the first two bands is plotted in  Fig. \ref{Pic1}. The ground state is chiral for type-I Dirac and Weyl spectra, whereas the ground state for type-II Dirac and Weyl spectra is independent of the chirality.\cite{RefI15}
\subsection*{The quantum theory of the planar Hall effect}
The energy density of states for tilted Dirac and Weyl semimetals in quantizing magnetic field is
\begin{equation}
\begin{split}
D(E)=\frac{g}{4\pi^{2}l_{B}^{2}v_{F}\hbar\gamma}\{1-\frac{\gamma v_{T}^{z}}{v_{F}}(1+n_{max}(n_{max}+1))\\
+2\sum_{n=1}^{n_{max}}\frac{E\gamma}{\sqrt{E^{2}\gamma^{2}-\frac{np_{B}^{2}v_{F}^{2}}{\gamma}|1-\gamma^{2}(\frac{v_{T}^{z}}{v_{F}})^{2}|}}\}.
\end{split}
\end{equation}
The above equation shows oscillations in the energy density of states with the changing magnetic field or the energy $E$, $E^{2}\gamma^{2}=v_{F}^{2}p_{B}^{2}n(1-\gamma^{2}(\frac{v_{T}^{z}}{v_{F}})^{2})$. The magnetoconductivity, and the magnetoresistivity also show these oscillations. The above formula for the energy density of states reproduces the result of Refs. [13,14] by turning off the tilt parameters$(\gamma=1,v_{T}^{x}=0,v_{T}^{z}=0)$ and setting $g=1$. 

\begin{figure}[t!]
	\centering
	\includegraphics[width=80mm]{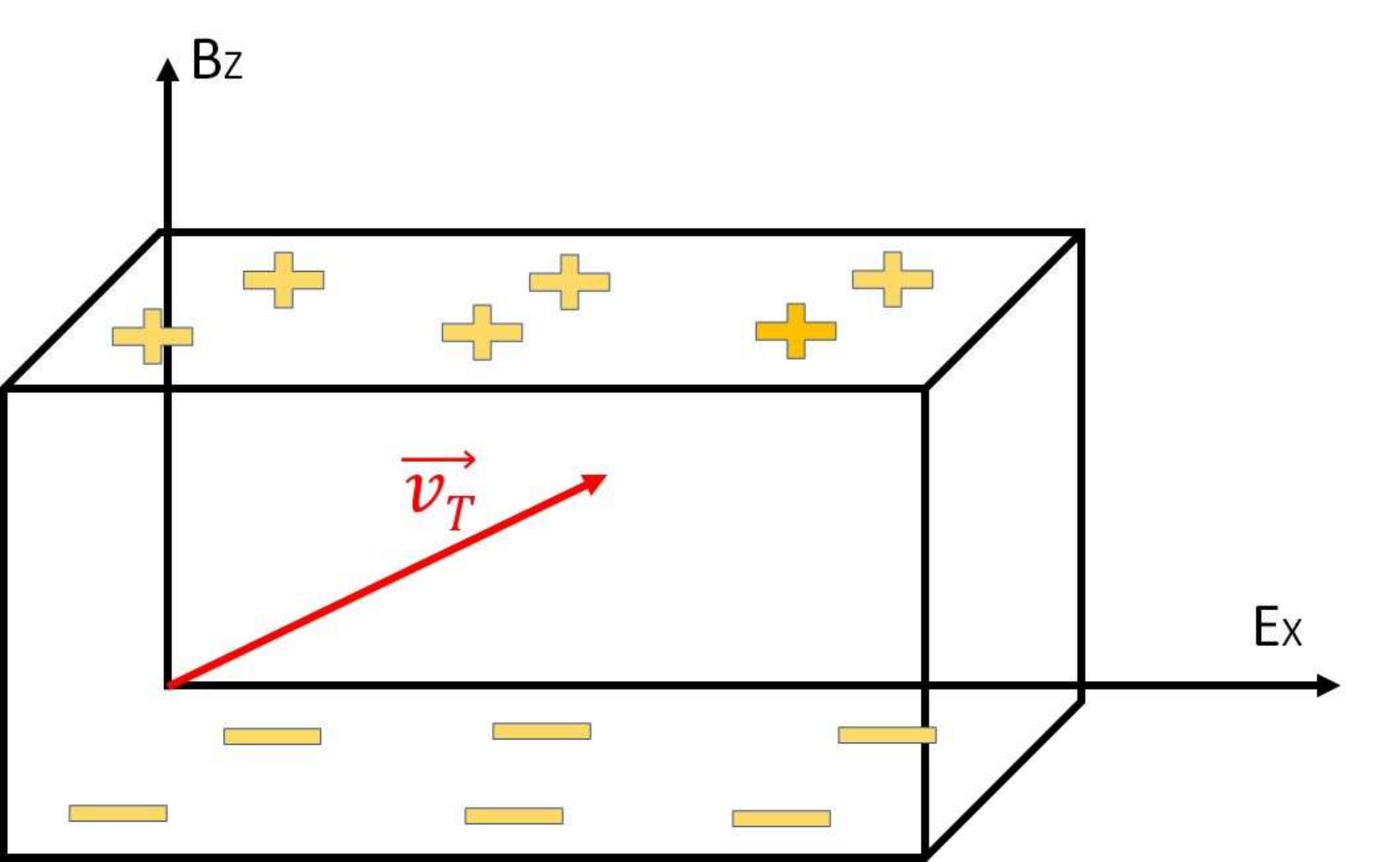}
	\caption{The schematic illustration of the planar Hall effect. \label{Pic2}}
\end{figure}

For completeness we derive the formulas for all the components of the magnetoconductivity and the magnetoresistivity tensors. We use the Kubo linear response theory for studying the quantum magnetotransport. The current current correlation function is given below.\cite{RefI35}
\begin{equation}
\begin{split}
Q_{ij}(\Omega)=2e^{2}T\sum_{\omega}\int\frac{dp_{z}}{2\pi\hbar}\int\frac{dp_{y}}{2\pi\hbar}\int dx'\\
Tr[\bar{v}_{i}G(p_{y},p_{z},\Omega+\omega,x,x')\bar{v}_{j}G(p_{y},p_{z},\omega,x',x)]
\end{split}
\end{equation}
In the above formula $\omega$ and $\Omega$ are Matsubara frequencies.
The retarded Greens function is
\begin{equation}
G^{r}(\omega,x,x',p_{z},p_{y})=\sum_{n,s}\frac{\psi^{s}_{n}(\vec{x})\psi^{s\dagger}_{n}(\vec{x'})}{\omega-\epsilon^{s}_{n}(p_{z})+i\Gamma}.
\end{equation}
Here $\Gamma=\frac{\hbar}{2\tau}$ is used for the finite broadening of the Landau levels.
The components of the velocity after the hyperbolic transformation are 
\begin{equation}
\bar{v}_{x}=\frac{1}{\gamma}v_{F}\sigma_{x},\quad\bar{v}_{y}=v_{F}\sigma_{y},\quad\bar{v}_{z}=\gamma v_{T}^{z}(\sigma_{0}+\frac{v_{T}^{x}}{v_{F}}\sigma_{x})+v_{F}\sigma_{z}.
\end{equation}
At this point we find it convenient to introduce a new tensor, $s_{ij}$, that only involves the Pauli matrices.
\begin{equation}
\begin{split}
s_{ij}(\Omega)=2e^{2}T\sum_{\omega}\int\frac{dp_{z}}{2\pi\hbar}\int\frac{dp_{y}}{2\pi\hbar}\int dx'\\
Tr[\sigma_{i}G(p_{y},p_{z},\Omega+\omega,x,x')\sigma_{j}G(p_{y},p_{z},\omega,x',x)]
\end{split}
\end{equation}
For finding static response we set $s_{ij}=lim_{\Omega\rightarrow0}\;\frac{s_{ij}(\Omega)}{i\Omega}$.
 The different elements of the tensor $s_{ij}$ are given below. 
\begin{equation}
\begin{split}
s_{xx}=\frac{e^{2}\beta \tau}{16\pi l_{B}^{2}}\sum_{n}\int\frac{dp_{z}}{2\pi\hbar}[\frac{\hbar^{2}}{\tau^{2}(\epsilon^{+}_{n}(p_{z})+\epsilon^{+}_{n-1}(p_{z}))^{2}+\hbar^{2}}\\
[a_{n}^{2}a_{n-1}^{2}(sech^{2}(\frac{\beta\epsilon^{-}_{n}(p_{z})}{2})+sech^{2}(\frac{\beta\epsilon^{+}_{n-1}(p_{z})}{2}))\\
+b_{n}^{2}b_{n-1}^{2}(sech^{2}(\frac{\beta\epsilon^{+}_{n}(p_{z})}{2})+sech^{2}(\frac{\beta\epsilon^{-}_{n-1}(p_{z})}{2}))]\\+\frac{\hbar^{2}}{\tau^{2}(\epsilon^{+}_{n}(p_{z})-\epsilon^{+}_{n-1}(p_{z}))^{2}+\hbar^{2}}\\
[a_{n}^{2}b_{n-1}^{2}((sech^{2}(\frac{\beta\epsilon^{-}_{n}(p_{z})}{2})+sech^{2}(\frac{\beta\epsilon^{-}_{n-1}(p_{z})}{2})))\\+a_{n-1}^{2}b_{n}^{2}(sech^{2}(\frac{\beta\epsilon^{+}_{n}(p_{z})}{2})+sech^{2}(\frac{\beta\epsilon^{+}_{n-1}(p_{z})}{2}))]]
\end{split}
\end{equation}
\begin{equation}
\begin{split}
s_{zz}=\frac{e^{2}\beta \tau}{16\pi l_{B}^{2}}\sum_{n}\int\frac{dp_{z}}{2\pi\hbar}[a_{n}^{4}+b_{n}^{4}\\
+2a_{n}^{2}b_{n}^{2}\frac{\hbar^{2} }{\tau^{2}(2\epsilon^{+}_{n}(p_{z}))^{2}+\hbar^{2}}]
\\
[sech^{2}(\frac{\beta\epsilon^{-}_{n}(p_{z})}{2})+sech^{2}(\frac{\beta\epsilon^{+}_{n}(p_{z})}{2})]
\end{split}
\end{equation} 
\begin{equation}
\begin{split}
s_{xy}=\frac{\hbar e^{2}}{4\pi l_{B}^{2}}\sum_{n}\int\frac{dp_{z}}{2\pi\hbar}[\frac{1}{(\epsilon^{+}_{n}(p_{z})+\epsilon^{+}_{n-1}(p_{z}))^{2}}(a_{n}^{2}a_{n-1}^{2}\\
[\tanh[\frac{\beta\epsilon^{-}_{n}(p_{z})}{2}]-\tanh[\frac{\beta\epsilon^{+}_{n-1}(p_{z})}{2}]]\\+b_{n}^{2}b_{n-1}^{2}[\tanh[\frac{\beta\epsilon^{+}_{n}(p_{z})}{2}]-\tanh[\frac{\beta\epsilon^{-}_{n-1}(p_{z})}{2}]])\\+\frac{1}{(\epsilon^{+}_{n}(p_{z})-\epsilon^{+}_{n-1}(p_{z}))^{2}}\\
(a_{n}^{2}b_{n-1}^{2}[\tanh[\frac{\beta\epsilon^{-}_{n}(p_{z})}{2}]-\tanh[\frac{\beta\epsilon^{-}_{n-1}(p_{z})}{2}]]\\+a_{n-1}^{2}b_{n}^{2}[\tanh[\frac{\beta\epsilon^{+}_{n}(p_{z})}{2}]-\tanh[\frac{\beta\epsilon^{+}_{n-1}(p_{z})}{2}]])]
\end{split}
\end{equation}
The remaining elements of the function $s_{ij}$ are found by exploiting symmetry of the Pauli matrices: $s_{ij}$, $s_{yy}=s_{xx}$, and $s_{yx}=-s_{xy}$.
All the elements of the conductivity matrix $\sigma_{ij}$ are found by using the $s_{ij}$ tensor.
\begin{equation}
\begin{split}
[\sigma_{ij}]=\begin{bmatrix}
\sigma_{xx} & \sigma_{xy} &\sigma_{xz}\\
\sigma_{yx} & \sigma_{yy} & \sigma_{yz}\\
\sigma_{zx} &\quad \sigma_{zy} &\sigma_{zz}
\end{bmatrix}
\end{split}
\end{equation} 
\begin{equation}
\sigma_{xx}=s_{xx}\frac{v_{F}^{2}}{\gamma^{2}},\quad \sigma_{xy}=s_{xy}\frac{v_{F}^{2}}{\gamma},\quad\sigma_{xz}=s_{xx}v_{T}^{x}v_{T}^{z} \label{Eq:19}
\end{equation}
\begin{equation}
\sigma_{yx}=-\sigma_{xy},\quad \sigma_{yy}=s_{yy}v_{F}^{2},\quad\sigma_{yz}=s_{yx}v_{T}^{x}v_{T}^{z}\gamma\label{Eq:20}
\end{equation}
\begin{equation}
\begin{split}
\sigma_{zx}=\sigma_{xz},\quad \sigma_{zy}=-\sigma_{yz},\\ \sigma_{zz}=s_{zz}(v_{F}^{2}+(v_{T}^{z})^{2}\gamma^{2})+s_{xx}(v_{T}^{z})^{2}\gamma^{2}(\frac{v_{T}^{x}}{v_{F}})^{2}\label{Eq:21}
\end{split}
\end{equation}
The new components of the conductivity matrix ($\sigma_{xz},\;\sigma_{zx},\;\sigma_{yz},\;\sigma_{zy}$) are directly proportional to the tilt velocity $v_{T}^{x},\;v_{T}^{z}$. In the absence of any one component of the tilt velocity ($v_{T}^{x}=0,\textrm{or}\;v_{T}^{z}=0$), the structure of the conductivity matrix does not show any signature of the tilted Dirac and Weyl spectra because it is the product $v_{T}^{x}v_{T}^{z}$ which enters into these components. The transverse conductivity $\sigma_{xx}$, and the Hall conductivity $\sigma_{xy}$ are renormalized by the Lorentz factor $\gamma$. The longitudinal conductivity $\sigma_{zz}$ is enhanced by the parallel component of the tilt velocity. The response of the transverse conductivity $s_{xx}$ is also added up with the longitudinal conductivity $\sigma_{zz}$. For comparison the equation of conductivity matrix in the absence of the perpendicular component of tilt velocity is    
\begin{equation}
\begin{split}
[\sigma_{ij}]=\begin{bmatrix}
s_{xx}v_{F}^{2} & s_{xy}v_{F}^{2} &0\\
s_{yx}v_{F}^{2} & s_{yy}v_{F}^{2} & 0\\
0 &\quad 0 &\quad s_{zz}(v_{F}^{2}+(v_{T}^{z})^{2})
\end{bmatrix}.
\end{split}
\end{equation} 
In general the resistivity matrix 
\begin{equation}
\begin{split}
[\rho_{ij}]=\begin{bmatrix}
\rho_{xx} & \rho_{xy} &\rho_{xz}\\
\rho_{yx} & \rho_{yy} & \rho_{yz}\\
\rho_{zx} & \rho_{zy} & \rho_{zz}
\end{bmatrix},
\end{split}
\end{equation}
and its different components are
\begin{equation}
\rho_{xx}=\frac{(s_{xx}^{2}+s_{xy}^{2})(v_{T}^{z})^{2}(\frac{v_{T}^{x}}{v_{F}})^{2}+s_{xx}s_{zz}((v_{T}^{z})^{2}+(\frac{v_{F}}{\gamma})^{2})}{(s_{xx}^{2}+s_{xy}^{2})s_{zz}(\frac{v_{F}}{\gamma})^{2}((v_{T}^{z})^{2}+(\frac{v_{F}}{\gamma})^{2})}\label{Eq:24}
\end{equation}
\begin{equation}
\rho_{yy}=\frac{1}{v^{2}_{F}}\frac{s_{xx}}{s^{2}_{xx}+s^{2}_{xy}}\label{Eq:25}
\end{equation}
\begin{equation}
\rho_{xy}=-\frac{s_{xy}\gamma}{(s_{xx}^{2}+s_{xy}^{2})v_{F}^{2}}\label{Eq:26}
\end{equation}
\begin{equation}
\rho_{xz}=-\frac{1}{s_{zz}}\frac{v_{T}^{z}v_{T}^{x}}{v_{F}^{2}((v_{T}^{z})^{2}+(\frac{v_{F}}{\gamma})^{2})}\label{Eq:27}
\end{equation}
\begin{equation}
\rho_{zz}=\frac{1}{s_{zz}}\frac{1}{\gamma^{2}((v_{T}^{z})^{2}+(\frac{v_{F}}{\gamma})^{2})}.\label{Eq:28}
\end{equation}
The remaining elements of the resistivity matrix are redundant by symmetry: $\rho_{yx}=-\rho_{xy},\;\;\rho_{zx}=\rho_{xz}$, $\rho_{yz}=\rho_{zy}=0$.
The transverse resistivity matrix $\rho_{xx}$ contains all elements of the tensor $s_{ij}$. The Hall and longitudinal resistivities, $\rho_{xy}$, and $\rho_{zz}$, are renormalized by the Lorentz factor $\gamma$ and the tilt velocity. 
In the above formula the presence of both components of the tilt velocity makes the planar Hall effect $\rho_{xz}$ finite. A schematic picture of the planar Hall effect in a slab of the 3D Dirac and Weyl gases is shown in the Fig. \ref{Pic2}. 
In general a sample of 3D Dirac and Weyl semimetals with tilted spectrum supports the planar Hall effect.  The tilt velocity can be set in the plane formed by the perpendicular electric and magnetic fields. This accumulates the Hall voltage in the plane of the electric and magnetic fields. The planar Hall voltage is present even in the absence of the magnetic field because with the tilt the energy spectrum is asymmetric in the $\hat{z}$ and $\hat{x}$ direction.
\subsection*{The Semiclassical theory of the planar Hall effect} 
For the semiclassical study of magnetotransport in the Dirac and Weyl semimetals we solve the Boltzmann equation in the relaxation time approximation 
\begin{equation}
\begin{split}
\frac{\partial f_{\chi}}{\partial t}+\vec{\dot{x}}\cdot\vec{\bigtriangledown}_{x}f_{\chi}+\frac{1}{\hbar}\vec{F}\cdot\vec{\bigtriangledown}_{k}f_{\chi}=-\frac{\delta f_{\chi}}{\tau_{tr}},\label{Eq:N}
\end{split}
\end{equation} 
where
\begin{equation}
\begin{split}
\vec{F}=\frac{1}{1+\frac{e}{\hbar}\vec{B}\cdot\vec{\Omega}}(e\vec{E}+e\vec{v}\times\vec{B}+\frac{e^{2}}{\hbar}(\vec{E}\cdot\vec{B})\vec{\Omega})\\
\vec{\dot{x}}=\frac{1}{1+\frac{e}{\hbar}\vec{B}\cdot\vec{\Omega}}(\vec{v}+\frac{e}{\hbar}\vec{E}\times\vec{\Omega}+\frac{e}{\hbar}(\vec{\Omega}\cdot\vec{v})\vec{B}).
\end{split}
\end{equation} 
$\vec{F}$ is the force acting on Dirac and Weyl fermions gas, $\vec{\dot{x}}$ is the group velocity, $\tau_{tr}$ is the transport time, and $\vec{\Omega}=-\chi\frac{\hat{k}}{2k^{2}}$ is the Berry curvature. The apparent difference in equations of velocity and force for the Dirac and Weyl gases enters due to the Berry curvature. For the study of planar Hall effect we ignore the effect of the Lorentz force\cite{RefI39} and assume a constant in time and space distribution function.
\begin{equation}
\delta f_{\chi}(x,k,t)\equiv \delta f^{\parallel}_{\chi}(k)
\end{equation}
 We use the linear order in electric field approximation for calculating current.
\begin{eqnarray}
\delta f^{\parallel}_{x}&=&A\tau_{tr}e(\vec{E}+\frac{e}{\hbar}(\vec{E}\cdot\vec{B})\vec{\Omega})\cdot\vec{v}(-\frac{\partial f_{eq}(\epsilon_{k,\chi})}{\partial \epsilon_{k,\chi}})\\
j_{z}&=&e\sum_{\chi}\int\frac{d^{3}k}{(2\pi)^{3}}A^{-1}\dot{x}_{z}\delta f^{\parallel}_{x}
\end{eqnarray} 
In the formula for the current we have included the phase space correction factor $A=\frac{1}{1+\frac{e}{\hbar}\vec{B}\cdot\vec{\Omega}}$. The planar Hall conductivity is
\begin{equation}
\begin{split}
\sigma_{zx}=e^{2}\tau_{tr}\sum_{\chi}\int\frac{d^{3}k}{(2\pi)^{3}}A(v_{z}+\frac{e}{\hbar}B_{z}(\vec{\Omega}\cdot\vec{v}))\\
(v_{x}+\frac{e}{\hbar}B_{x}(\vec{\Omega}\cdot\vec{v}))(-\frac{\partial f_{eq}(\epsilon_{k,\chi})}{\partial \epsilon_{k,\chi}}).\\
\end{split}
\end{equation} 
The energy dispersion and the velocities of the tilted Dirac and Weyl semimetals are
\begin{equation}
\begin{split}
\epsilon_{k,\chi}=\chi(\hbar k_{x}v_{T}^{x}+\hbar k_{z}v_{T}^{z}+s\hbar|k|v_{F}),\\
\end{split}
\end{equation} 
\begin{equation}
\begin{split}
v_{x}=\chi(v_{T}^{x}+sv_{F}\cos\phi\sin\theta)\\
v_{y}=\chi sv_{F}\sin\phi\sin\theta\\
v_{z}=\chi(v_{T}^{z}+sv_{F}\cos\theta).
\end{split}
\end{equation} 
The sum of the particles densities ($N^{+}$, $N^{-}$) and the currents ($J^{+}$, $J^{-}$) are conserved in the presence of the chiral anomaly and the tilt velocity.
\begin{equation}
\begin{split}
\sum_{\chi=\pm}\int \frac{d^{3}k}{2\pi^{3}}A^{-1}\frac{\delta f_{\chi}^{\parallel}}{\tau_{tr}}=e\sum_{\chi=\pm}\chi\int \frac{d^{3}k}{2\pi^{3}}\frac{\partial f_{eq}(\epsilon_{k,\chi})}{\partial\epsilon_{k,\chi}}\\
[(v_{T}^{x}E_{x}+v_{T}^{z}E_{z})
+\frac{e}{\hbar}(\vec{E}\cdot\vec{B})(\vec{\Omega}\cdot\vec{v})]=0
\end{split}
\end{equation}
\begin{equation}
\frac{\partial}{\partial t}(N^{+}+N^{-})+\vec{\nabla}\cdot(\vec{J}^{+}+\vec{J}^{-})=0
\end{equation}
However, the chiral anomaly and tilt velocity creates imbalance between population densities of different chirality particles
\begin{equation}
\delta N=N^{+}-N^{-}=\int \frac{d^{3}k}{2\pi^{3}}A^{-1}(\delta f_{+}^{\parallel}-\delta f_{-}^{\parallel}).
\end{equation}
In the absence of a magnetic field this imbalance is created by the tilt velocity
\begin{equation}
\begin{split}
\delta N=N^{+}-N^{-}=-e\tau_{tr}(v_{T}^{x}E_{x}+v_{T}^{z}E_{z})\\
\int \frac{d^{3}k}{2\pi^{3}}(\frac{\partial f_{eq}(\epsilon_{k,+})}{\partial\epsilon_{k,+}}+\frac{\partial f_{eq}(\epsilon_{k,-})}{\partial\epsilon_{k,-}}).
\end{split}
\end{equation}
This imbalance in turn creates planar Hall effect 
\begin{eqnarray}
\sigma_{zx}&=&e^{2}\tau_{tr}v_{T}^{x}v_{T}^{z}\int\frac{d^{3}k}{(2\pi)^{3}}(-\frac{\partial f_{eq}(\epsilon_{k})}{\partial \epsilon_{k}})\\
\frac{\partial f_{eq}(\epsilon_{k})}{\partial \epsilon_{k}}&=&(\frac{\partial f_{eq}(\epsilon_{k,+})}{\partial \epsilon_{k,+}}+\frac{\partial f_{eq}(\epsilon_{k,-})}{\partial \epsilon_{k,-}}).
\end{eqnarray}
The planar Hall effect arising due to chiral anomaly changes its functional dependence on magnetic field in the presence of tilt velocity, $\rho_{xz}\sim B_{z}$,\cite{RefI36} 
since the planar Hall effect without tilt velocity is proportional to $B_{x}B_{z}$. Because we have allowed for both $v_{T}^{x}$ and $v_{T}^{z}$ to be non zero, the planar Hall effect derived here exists even for zero magnetic field. For magnetic field along one axis like $B_{z}$ the planar Hall effect due to chiral anomaly will vanish. Therefore the tilt velocity provides another fingerprint to detect chiral anomaly via the planar Hall effect.
\section*{Results and Discussions}  
In the region of strong magnetic field when the Landau levels are well resolved, the spacing between the two adjacent Landau levels $\Delta E=\frac{\sqrt{2}\hbar v_{F}}{l_{B}}\sim 10 \;meV$ is much greater than the thermal energy $\frac{k_{B}T}{\Delta E}\sim0.01 $ and the impurity broadening $\frac{\Gamma}{\Delta E}\sim 0.01 $. In our calculations for the current-current correlation function we have assumed the quasiparticle lifetime $\tau_{q}$ is of the same order as the transport lifetime $\tau_{tr}\sim\tau_{q}$. In general the calculations of the transport time requires the self-consistent Born approximation, and this can differ by order of magnitudes with the quasiparticle lifetime. This introduces the significant new physics by updating the quasiparticle lifetime with the transport time. However, in the domain of a strong magnetic field where the Landau levels spacing is the largest magnetotransport parameter, the transport and the quasiparticle lifetimes are reasonably of the same order.\cite{RefI16} The case of overlapping Landau levels and intermediate temperature region, where the interaction corrections are important, is discussed in the Refs. [13, 14].

Experimentally the quasiparticle lifetime, Fermi velocity and the effective mass is measured by fitting the De Haas Van Alphen oscillations or the Shubnikov-de Haas oscillations with the Liftshitz-Kosevich formula. By using the experimental data of $TaP$ \cite{RefI25} for the Fermi velocity $v_{F}\sim0.125\; (eV-nm)$, Fermi energy $E_{F}\sim0.039\;eV$, impurity broadening $\Gamma\sim0.0023\;eV$, and the mobility $\mu_{Q}\sim0.36\;\frac{m^{2}}{Vs}$, we found the quasiparticle lifetime $\tau_{q}\sim2.8\times10^{-13}s$, and the transport time $\tau_{tr}\sim4\times10^{-13}s$ are of the same order $\tau_{q}\sim\tau_{tr}$.   

\begin{figure}[t!]
	\centering
	\includegraphics[width=80mm]{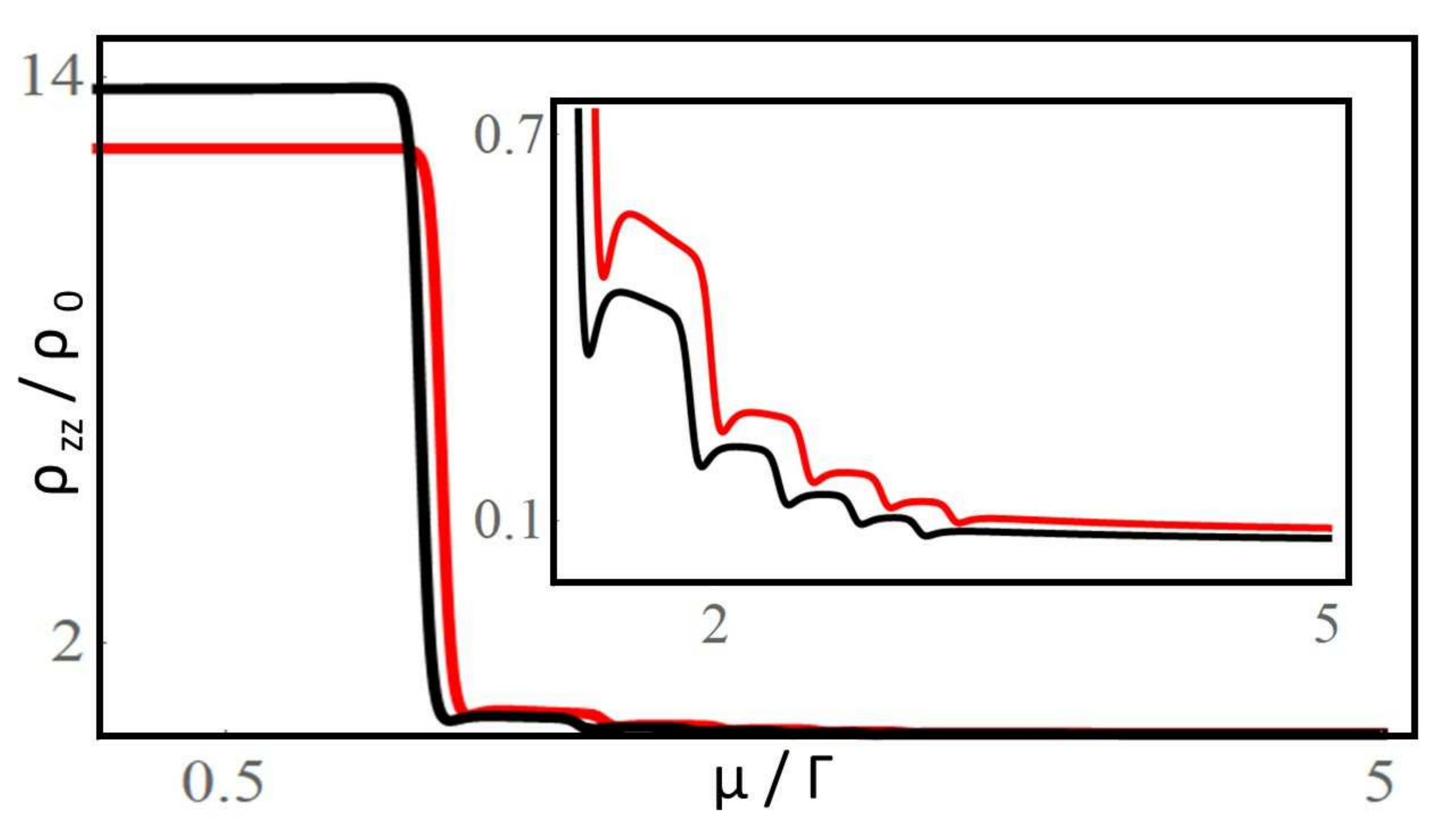}
	\caption{The longitudinal resistivity $\rho_{zz}$. The solid black curve is the plot for no tilt in the Dirac and Weyl spectrum, whereas the solid red curve is the plot in the presence of the tilt parameters, $\frac{v_{T}^{z}}{v_{F}}=0.2$, and $\frac{v_{T}^{x}}{v_{F}}=0.2$.  Here $\rho_{0}=\frac{32\pi^{2}l^{2}_{B}}{e^{2}v_{F}\beta}$, $\beta=100\;meV^{-1}$, and $\frac{\sqrt{2}\hbar v_{F}}{l_{B}}=1\;meV$.  \label{Pic3}}
\end{figure}
\begin{figure}[b!]
	\centering
	\includegraphics[width=80mm]{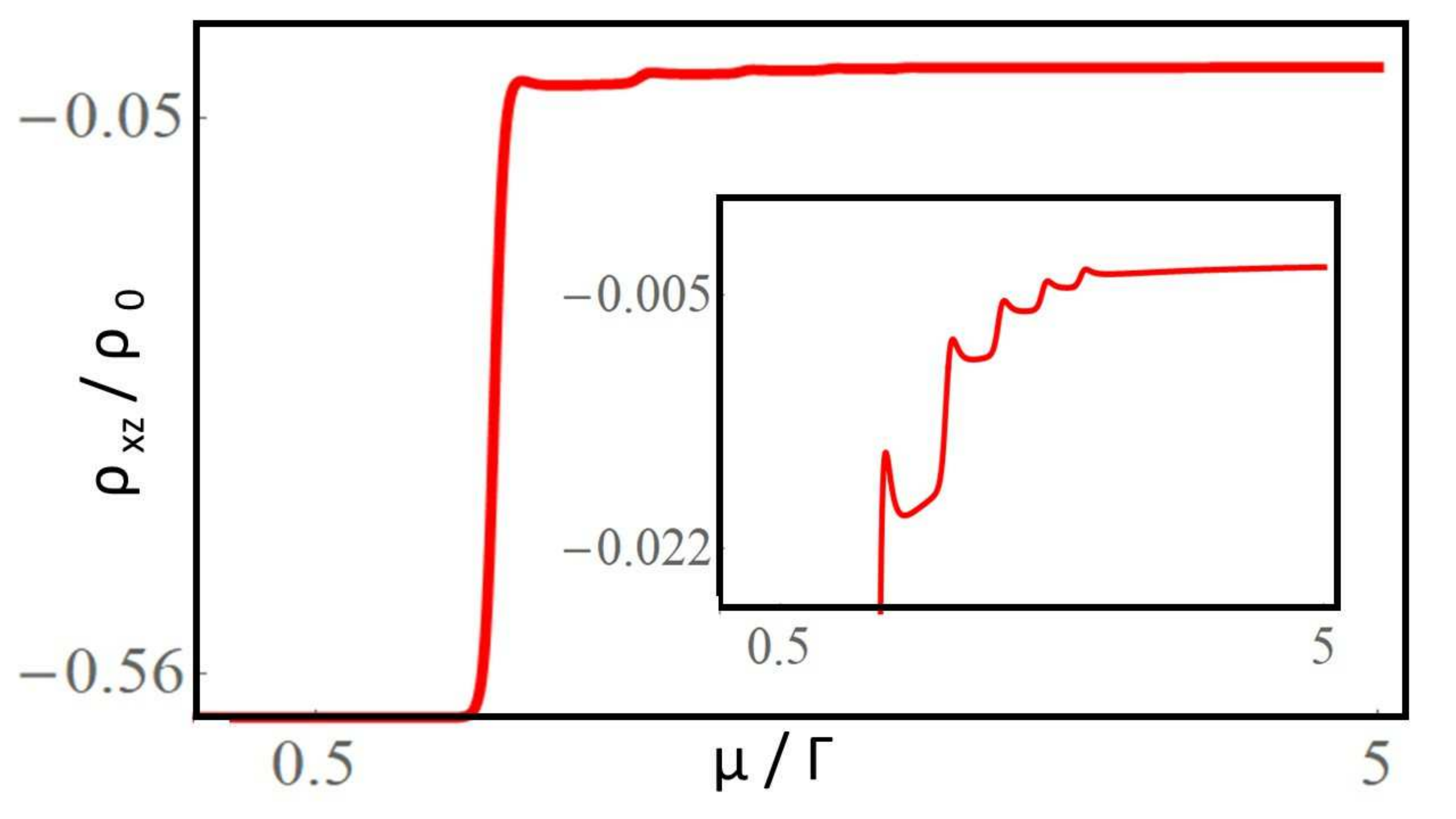}
	\caption{The planar Hall resistivity $\rho_{xz}$. The parameters used here are the same as used in the $\rho_{zz}$. \label{Pic4}}
\end{figure}
The longitudinal and planar Hall resistivities $\rho_{zz},\;\rho_{xz}$ are plotted in the Figs. \ref{Pic3} and \ref{Pic4}. The oscillations in the longitudinal resistivity is also shown in the Ref. [15,20]. The oscillations in the longitudinal resistivity is a unique feature of the 3D Dirac and Weyl semimetals. 
The planar Hall resistivity enters into the magnetoresistivity tensor due to the tilt velocity $v_{T}^{x}\;\textrm{and}\;v_{T}^{z}$. The planar Hall resistivity is directly proportional to the longitudinal resistivity $\rho_{xz}=-\frac{v_{T}^{x}v_{T}^{z}\gamma^{2}}{v_{F}^{2}}\rho_{zz}$. Therefore, the oscillations in the planar Hall resistivity provide another probe for the detection of the chiral anomaly. The planar Hall effect can be detected in materials like $TaP$,\cite{RefI25} and $WTe_{2}$.\cite{RefI26} By allowing the tilt velocity to make a finite angle in the plane formed by the perpendicular electric and magnetic fields, the planar Hall effect and the Shubnikov de Haas oscillations can be detected. 

\section*{Conclusions}
In this study we have investigated the role of tilt velocity in magnetotransport properties. We have shown that
all the components of the magnetoconductivity and the magnetoresisitivity matrix are modified by the tilt velocity. The longitudinal $\rho_{zz}$, transverse $\rho_{xx}$, Hall $\rho_{xy}$, and planar Hall $\rho_{xz}$ magnetoresistivities resonate whenever the chemical potential crosses the energy band $\mu=\epsilon_{n}^{s}$. The tilt velocity shifts the position and renormalizes the weight of these peaks. This also mixes the different elements of the resistivity matrix. The planar Hall effect is directly proportional to the tilt velocity. 

The planar Hall effect can become fingerprint to detect nontrivial magnetotransport in the 3D Dirac and Weyl gases with the tilted spectrum. The planar Hall effect is directly proportional to the longitudinal resistivity, which oscillates with magnetic field.\cite{RefI24} The oscillations in the planar Hall effect can identify a material that support the tilted Dirac and Weyl spectra with the unique anomalous transport features.   


\newpage
\begin{thebibliography}{99}  
	\bibitem{RefI1}Q. L. He, L. Pan, A. L. Stern, E. C. Burks, X. Che, G. Yin, J. Wang, B. Lian, Q. Zhou, E. S. Choi, K. Murata, X. Kou, Z. Chen, T. Nie, Q. Shao, Y. Fan, S. C. Zhang, K. Liu, J. Xia, and K. L. Wang, Science 357, 294 (2017).  
	\bibitem{RefI2}S. Y. Xu, I. Belopolski, N. Alidoust, M. Neupane, G. Bian, C. Zhang, R. Sankar, G. Chang, Z. Yuan, C. C. Lee, S. M. Huang, H. Zheng, J. Ma, D. S. Sanchez, B. Wang, A. Bansil, F. Chou, P. P. Shibayev, H. Lin, S. Jia, and M. Z. Hasan, Science 349, 613 (2015). 
	\bibitem{RefI3}B. Q. Lv, H. M. Weng, B. B. Fu, X. P. Wang, H. Miao, J. Ma, P. Richard, X. C. Huang, L. X. Zhao, G. F. Chen, Z. Fang, X. Dai, T. Qian, and H. Ding, Phys. Rev. X 5, 031013 (2015).   
	\bibitem{RefI4}S. Y. Xu, N. Alidoust, G. Chang, H. Lu, B. Singh, I. Belopolski, D. Sanchez, X. Zhang, G. Bian, H. Zheng, M. A. Husanu, Y. Bian, S. M. Huang, C. H. Hsu, T. R. Chang, H. T. Jeng, A. Bansil, V. N. Strocov, H. Lin, S. Jia, and M. Z. Hasan, Sci. Adv. 3, e1603266 (2017).   
	\bibitem{RefI5}A. A. Soluyanov, D. Gresch, Z. Wang, Q. Wu, M. Troyer, X. Dai, and B. A. Bernevig, Nature (London) 527, 495 (2015).  
	\bibitem{RefI6}Z. Wang, D. Gresch, A. A. Soluyanov, W. Xie, S. Kushwaha, X. Dai, M. Troyer, R. J. Cava, and B. A. Bernevig, Phys. Rev. Lett. 117, 056805 (2016).
	\bibitem{RefI12}N. Kumar, C. Felser, and C. Shekhar, Phys. Rev. B 98, 041103 (2018).   
	\bibitem{RefI15}S. Tchoumakov, M. Civelli, and M. O. Goerbig
	Phys. Rev. Lett. 117 086402 (2016) 
	\bibitem{RefI32}A. A. Burkov and L. Balents, Phys. Rev. Lett. 107, 127205 (2011). 
	\bibitem{RefI33}S. Nandy, G. Sharma, A. Taraphder, and S. Tewari, Phys. Rev. Lett. 119, 176804 (2017). 
	\bibitem{RefI34}A. A. Burkov, Phys. Rev. B 96, 041110 (2017).     
	\bibitem{RefI36}Gen Yin, Jie-Xiang Yu, Yizhou Liu, Roger K. Lake, Jiadong Zang, and Kang L. Wang, Phys. Rev. Lett. 122, 106602 (2019).
	\bibitem{RefI16}J. Klier, I. V. Gornyi, and A. D. Mirlin, Phys. Rev. B 92, 205113 (2015). 
	\bibitem {RefI17}J. Klier, I. V. Gornyi, and A. D. Mirlin , Phys. Rev. A. 96, 214209 (2017).  
	\bibitem{RefI19}M. Trescher, B. Sbierski, P. W. Brouwer, and E. J. Bergholtz, Phys. Rev. B 95, 045139 (2017).
	\bibitem{RefI24}M. X. Deng, G. Y. Qi, R. Ma, R. Shen, R. Q. Wang, L. Sheng, and D. Y. Xing, Phys. Rev. Lett. 122, 036601 (2019).
	\bibitem{RefI25}C. L. Zhang, S. Y. Xu, C. M. Wang, Z. Lin, Z. Z. Du, C. Guo,
	C. C. Lee, H. Lu, Y. Feng, S. M. Huang, G. Chang,
	C. H. Hsu, H. Liu, H. Lin, L. Li, C. Zhang, J. Zhang, X. C. Xie,
	T. Neupert, M. Z. Hasan, H. Z. Lu, J. Wang, and S. Jia, Nat. Phys. 13, 979 (2017). 
	\bibitem{RefI26}P. Li, Y. Wen, X. He, Q. Zhang, C. Xia, Z. M. Yu, S. A. Yang, Z. Zhu, H. N. Alshareef, and X. X. Zhang, 
	 Nat. Commun. 8, 2150 (2017).  
	 \bibitem{RefI35}X. Xiao, K. T. Law, and P. A. Lee, Phys. Rev. B 96, 165101 (2017).  
	 \bibitem{RefI36}Da Ma, Hua Jiang, Haiwen Liu, and X. C. Xie, Phys. Rev. B 99, 115121 (2019).  
	 \bibitem{RefI37}E. V. Gorbar, V. A. Miransky, I. A. Shovkovy, and P. O. Sukhachov, Low Temperature Physics 44, 635 (2018).
	\bibitem{RefI38}Qianqian Liu, Fucong Fei, Bo Chen, Xiangyan Bo, Boyuan Wei, Shuai Zhang, Minhao Zhang, Faji Xie, Muhammad Naveed, Xiangang Wan, Fengqi Song, and Baigeng Wang, Phys. Rev. B 99, 115119 (2019).
	\bibitem{RefI39}Aurélie Collaudin, Benoît Fauqué, Yuki Fuseya, Woun Kang, and Kamran Behnia, Phys. Rev. X 5, 021022 (2015).
	

	
	
	
	
\end{thebibliography}
\end{document}